\documentclass[conference,10pt]{IEEEtran}
\usepackage{balance}

\usepackage{cite}
\usepackage{graphicx}
\usepackage{amssymb}
\usepackage{amsmath}
\usepackage{amsthm}

\usepackage{paralist}

\usepackage{fancyref}
\usepackage{framed}
\usepackage{afterpage}
\usepackage{setspace}
\usepackage{microtype}

\usepackage{url}
\usepackage{booktabs}
\usepackage{algorithm}
\usepackage{algorithmic}
\usepackage{subfigure}
\usepackage{hyperref}
\usepackage{balance}

\usepackage{xcolor}
\usepackage{setspace}

\newcommand{\revision}[1]{#1}

\usepackage{amssymb}
\usepackage{amsfonts}
\usepackage{mathrsfs}
\usepackage{xspace}
\usepackage{bm}
\usepackage{upgreek}

\newcommand{\safemath}[2]{\newcommand{#1}{\ensuremath{#2}\xspace}}

\safemath{\bma}{\mathbf{a}}
\safemath{\bmb}{\mathbf{b}}
\safemath{\bmc}{\mathbf{c}}
\safemath{\bmd}{\mathbf{d}}
\safemath{\bme}{\mathbf{e}}
\safemath{\bmf}{\mathbf{f}}
\safemath{\bmg}{\mathbf{g}}
\safemath{\bmh}{\mathbf{h}}
\safemath{\bmi}{\mathbf{i}}
\safemath{\bmj}{\mathbf{j}}
\safemath{\bmk}{\mathbf{k}}
\safemath{\bml}{\mathbf{l}}
\safemath{\bmm}{\mathbf{m}}
\safemath{\bmn}{\mathbf{n}}
\safemath{\bmo}{\mathbf{o}}
\safemath{\bmp}{\mathbf{p}}
\safemath{\bmq}{\mathbf{q}}
\safemath{\bmr}{\mathbf{r}}
\safemath{\bms}{\mathbf{s}}
\safemath{\bmt}{\mathbf{t}}
\safemath{\bmu}{\mathbf{u}}
\safemath{\bmv}{\mathbf{v}}
\safemath{\bmw}{\mathbf{w}}
\safemath{\bmx}{\mathbf{x}}
\safemath{\bmy}{\mathbf{y}}
\safemath{\bmz}{\mathbf{z}}
\safemath{\bmzero}{\mathbf{0}}
\safemath{\bmone}{\mathbf{1}}

\bmdefine{\biad}{a}
\bmdefine{\bibd}{b}
\bmdefine{\bicd}{c}
\bmdefine{\bidd}{d}
\bmdefine{\bied}{e}
\bmdefine{\bifd}{f}
\bmdefine{\bigd}{g}
\bmdefine{\bihd}{h}
\bmdefine{\biid}{i}
\bmdefine{\bijd}{j}
\bmdefine{\bikd}{k}
\bmdefine{\bild}{l}
\bmdefine{\bimd}{m}
\bmdefine{\bind}{n}
\bmdefine{\biod}{o}
\bmdefine{\bipd}{p}
\bmdefine{\biqd}{q}
\bmdefine{\bird}{r}
\bmdefine{\bisd}{s}
\bmdefine{\bitd}{t}
\bmdefine{\biud}{u}
\bmdefine{\bivd}{v}
\bmdefine{\biwd}{w}
\bmdefine{\bixd}{x}
\bmdefine{\biyd}{y}
\bmdefine{\bizd}{z}

\bmdefine{\bixid}{\xi}
\bmdefine{\bilambdad}{\lambda}
\bmdefine{\bimud}{\mu}
\bmdefine{\bithetad}{\theta}
\bmdefine{\biphid}{\phi}
\bmdefine{\bideltad}{\delta}

\safemath{\bmia}{\biad}
\safemath{\bmib}{\bibd}
\safemath{\bmic}{\bicd}
\safemath{\bmid}{\bidd}
\safemath{\bmie}{\bied}
\safemath{\bmif}{\bifd}
\safemath{\bmig}{\bigd}
\safemath{\bmih}{\bihd}
\safemath{\bmii}{\biid}
\safemath{\bmij}{\bijd}
\safemath{\bmik}{\bikd}
\safemath{\bmil}{\bild}
\safemath{\bmim}{\bimd}
\safemath{\bmin}{\bind}
\safemath{\bmio}{\biod}
\safemath{\bmip}{\bipd}
\safemath{\bmiq}{\biqd}
\safemath{\bmir}{\bird}
\safemath{\bmis}{\bisd}
\safemath{\bmit}{\bitd}
\safemath{\bmiu}{\biud}
\safemath{\bmiv}{\bivd}
\safemath{\bmiw}{\biwd}
\safemath{\bmix}{\bixd}
\safemath{\bmiy}{\biyd}
\safemath{\bmiz}{\bizd}

\safemath{\bmxi}{\bixid}
\safemath{\bmlambda}{\bilambdad}
\safemath{\bmmu}{\bimud}
\safemath{\bmtheta}{\bithetad}
\safemath{\bmphi}{\biphid}
\safemath{\bmdelta}{\bideltad}

\safemath{\bA}{\mathbf{A}}
\safemath{\bB}{\mathbf{B}}
\safemath{\bC}{\mathbf{C}}
\safemath{\bD}{\mathbf{D}}
\safemath{\bE}{\mathbf{E}}
\safemath{\bF}{\mathbf{F}}
\safemath{\bG}{\mathbf{G}}
\safemath{\bH}{\mathbf{H}}
\safemath{\bI}{\mathbf{I}}
\safemath{\bJ}{\mathbf{J}}
\safemath{\bK}{\mathbf{K}}
\safemath{\bL}{\mathbf{L}}
\safemath{\bM}{\mathbf{M}}
\safemath{\bN}{\mathbf{N}}
\safemath{\bO}{\mathbf{O}}
\safemath{\bP}{\mathbf{P}}
\safemath{\bQ}{\mathbf{Q}}
\safemath{\bR}{\mathbf{R}}
\safemath{\bS}{\mathbf{S}}
\safemath{\bT}{\mathbf{T}}
\safemath{\bU}{\mathbf{U}}
\safemath{\bV}{\mathbf{V}}
\safemath{\bW}{\mathbf{W}}
\safemath{\bX}{\mathbf{X}}
\safemath{\bY}{\mathbf{Y}}
\safemath{\bZ}{\mathbf{Z}}

\safemath{\bZero}{\mathbf{0}}
\safemath{\bOne}{\mathbf{1}}
\safemath{\bDelta}{\mathbf{\Delta}}
\safemath{\bLambda}{\mathbf{\UpLambda}}
\safemath{\bPhi}{\mathbf{\Upphi}}
\safemath{\bSigma}{\mathbf{\Upsigma}}
\safemath{\bOmega}{\mathbf{\Upomega}}
\safemath{\bTheta}{\mathbf{\Uptheta}}

\bmdefine{\biAd}{A}
\bmdefine{\biBd}{B}
\bmdefine{\biCd}{C}
\bmdefine{\biDd}{D}
\bmdefine{\biEd}{E}
\bmdefine{\biFd}{F}
\bmdefine{\biGd}{G}
\bmdefine{\biHd}{H}
\bmdefine{\biId}{I}
\bmdefine{\biJd}{J}
\bmdefine{\biKd}{K}
\bmdefine{\biLd}{L}
\bmdefine{\biMd}{M}
\bmdefine{\biNd}{N}
\bmdefine{\biOd}{O}
\bmdefine{\biPd}{P}
\bmdefine{\biQd}{Q}
\bmdefine{\biRd}{R}
\bmdefine{\biSd}{S}
\bmdefine{\biTd}{T}
\bmdefine{\biUd}{U}
\bmdefine{\biVd}{V}
\bmdefine{\biWd}{W}
\bmdefine{\biXd}{X}
\bmdefine{\biYd}{Y}
\bmdefine{\biZd}{Z}

\bmdefine{\biDelta}{\Delta}
\bmdefine{\biLambda}{\Lambda}
\bmdefine{\biPhi}{\Phi}
\bmdefine{\biSigma}{\Sigma}
\bmdefine{\biOmega}{\Omega}
\bmdefine{\biTheta}{\Theta}

\safemath{\bimA}{\biAd}
\safemath{\bimB}{\biBd}
\safemath{\bimC}{\biCd}
\safemath{\bimD}{\biDd}
\safemath{\bimE}{\biEd}
\safemath{\bimF}{\biFd}
\safemath{\bimG}{\biGd}
\safemath{\bimH}{\biHd}
\safemath{\bimI}{\biId}
\safemath{\bimJ}{\biJd}
\safemath{\bimK}{\biKd}
\safemath{\bimL}{\biLd}
\safemath{\bimM}{\biMd}
\safemath{\bimN}{\biNd}
\safemath{\bimO}{\biOd}
\safemath{\bimP}{\biPd}
\safemath{\bimQ}{\biQd}
\safemath{\bimR}{\biRd}
\safemath{\bimS}{\biSd}
\safemath{\bimT}{\biTd}
\safemath{\bimU}{\biUd}
\safemath{\bimV}{\biVd}
\safemath{\bimW}{\biWd}
\safemath{\bimX}{\biXd}
\safemath{\bimY}{\biYd}
\safemath{\bimZ}{\biZd}

\safemath{\bimDelta}{\biDelta}
\safemath{\bimLambda}{\biLambda}
\safemath{\bimPhi}{\biPhi}
\safemath{\bimSigma}{\biSigma}
\safemath{\bimOmega}{\biOmega}
\safemath{\bimTheta}{\biTheta}

\safemath{\setA}{\mathcal{A}}
\safemath{\setB}{\mathcal{B}}
\safemath{\setC}{\mathcal{C}}
\safemath{\setD}{\mathcal{D}}
\safemath{\setE}{\mathcal{E}}
\safemath{\setF}{\mathcal{F}}
\safemath{\setG}{\mathcal{G}}
\safemath{\setH}{\mathcal{H}}
\safemath{\setI}{\mathcal{I}}
\safemath{\setJ}{\mathcal{J}}
\safemath{\setK}{\mathcal{K}}
\safemath{\setL}{\mathcal{L}}
\safemath{\setM}{\mathcal{M}}
\safemath{\setN}{\mathcal{N}}
\safemath{\setO}{\mathcal{O}}
\safemath{\setP}{\mathcal{P}}
\safemath{\setQ}{\mathcal{Q}}
\safemath{\setR}{\mathcal{R}}
\safemath{\setS}{\mathcal{S}}
\safemath{\setT}{\mathcal{T}}
\safemath{\setU}{\mathcal{U}}
\safemath{\setV}{\mathcal{V}}
\safemath{\setW}{\mathcal{W}}
\safemath{\setX}{\mathcal{X}}
\safemath{\setY}{\mathcal{Y}}
\safemath{\setZ}{\mathcal{Z}}
\safemath{\emptySet}{\varnothing}

\safemath{\colA}{\mathscr{A}}
\safemath{\colB}{\mathscr{B}}
\safemath{\colC}{\mathscr{C}}
\safemath{\colD}{\mathscr{D}}
\safemath{\colE}{\mathscr{E}}
\safemath{\colF}{\mathscr{F}}
\safemath{\colG}{\mathscr{G}}
\safemath{\colH}{\mathscr{H}}
\safemath{\colI}{\mathscr{I}}
\safemath{\colJ}{\mathscr{J}}
\safemath{\colK}{\mathscr{K}}
\safemath{\colL}{\mathscr{L}}
\safemath{\colM}{\mathscr{M}}
\safemath{\colN}{\mathscr{N}}
\safemath{\colO}{\mathscr{O}}
\safemath{\colP}{\mathscr{P}}
\safemath{\colQ}{\mathscr{Q}}
\safemath{\colR}{\mathscr{R}}
\safemath{\colS}{\mathscr{S}}
\safemath{\colT}{\mathscr{T}}
\safemath{\colU}{\mathscr{U}}
\safemath{\colV}{\mathscr{V}}
\safemath{\colW}{\mathscr{W}}
\safemath{\colX}{\mathscr{X}}
\safemath{\colY}{\mathscr{Y}}
\safemath{\colZ}{\mathscr{Z}}

\safemath{\opA}{\mathbb{A}}
\safemath{\opB}{\mathbb{B}}
\safemath{\opC}{\mathbb{C}}
\safemath{\opD}{\mathbb{D}}
\safemath{\opE}{\mathbb{E}}
\safemath{\opF}{\mathbb{F}}
\safemath{\opG}{\mathbb{G}}
\safemath{\opH}{\mathbb{H}}
\safemath{\opI}{\mathbb{I}}
\safemath{\opJ}{\mathbb{J}}
\safemath{\opK}{\mathbb{K}}
\safemath{\opL}{\mathbb{L}}
\safemath{\opM}{\mathbb{M}}
\safemath{\opN}{\mathbb{N}}
\safemath{\opO}{\mathbb{O}}
\safemath{\opP}{\mathbb{P}}
\safemath{\opQ}{\mathbb{Q}}
\safemath{\opR}{\mathbb{R}}
\safemath{\opS}{\mathbb{S}}
\safemath{\opT}{\mathbb{T}}
\safemath{\opU}{\mathbb{U}}
\safemath{\opV}{\mathbb{V}}
\safemath{\opW}{\mathbb{W}}
\safemath{\opX}{\mathbb{X}}
\safemath{\opY}{\mathbb{Y}}
\safemath{\opZ}{\mathbb{Z}}
\safemath{\opZero}{\mathbb{O}}
\safemath{\identityop}{\opI}

\safemath{\veca}{\bma}
\safemath{\vecb}{\bmb}
\safemath{\vecc}{\bmc}
\safemath{\vecd}{\bmd}
\safemath{\vece}{\bme}
\safemath{\vecf}{\bmf}
\safemath{\vecg}{\bmg}
\safemath{\vech}{\bmh}
\safemath{\veci}{\bmi}
\safemath{\vecj}{\bmj}
\safemath{\veck}{\bmk}
\safemath{\vecl}{\bml}
\safemath{\vecm}{\bmm}
\safemath{\vecn}{\bmn}
\safemath{\veco}{\bmo}
\safemath{\vecp}{\bmp}
\safemath{\vecq}{\bmq}
\safemath{\vecr}{\bmr}
\safemath{\vecs}{\bms}
\safemath{\vect}{\bmt}
\safemath{\vecu}{\bmu}
\safemath{\vecv}{\bmv}
\safemath{\vecw}{\bmw}
\safemath{\vecx}{\bmx}
\safemath{\vecy}{\bmy}
\safemath{\vecz}{\bmz}

\safemath{\veczero}{\bmzero}
\safemath{\vecone}{\bmone}
\safemath{\vecxi}{\bmxi}
\safemath{\veclambda}{\bmlambda}
\safemath{\vecmu}{\bmmu}
\safemath{\vectheta}{\bmtheta}
\safemath{\vecphi}{\bmphi}
\safemath{\vecdelta}{\bmdelta}

\safemath{\matA}{\bA}
\safemath{\matB}{\bB}
\safemath{\matC}{\bC}
\safemath{\matD}{\bD}
\safemath{\matE}{\bE}
\safemath{\matF}{\bF}
\safemath{\matG}{\bG}
\safemath{\matH}{\bH}
\safemath{\matI}{\bI}
\safemath{\matJ}{\bJ}
\safemath{\matK}{\bK}
\safemath{\matL}{\bL}
\safemath{\matM}{\bM}
\safemath{\matN}{\bN}
\safemath{\matO}{\bO}
\safemath{\matP}{\bP}
\safemath{\matQ}{\bQ}
\safemath{\matR}{\bR}
\safemath{\matS}{\bS}
\safemath{\matT}{\bT}
\safemath{\matU}{\bU}
\safemath{\matV}{\bV}
\safemath{\matW}{\bW}
\safemath{\matX}{\bX}
\safemath{\matY}{\bY}
\safemath{\matZ}{\bZ}
\safemath{\matzero}{\bmzero}

\safemath{\matDelta}{\bDelta}
\safemath{\matLambda}{\bLambda}
\safemath{\matPhi}{\bPhi}
\safemath{\matSigma}{\bSigma}
\safemath{\matOmega}{\bOmega}
\safemath{\matTheta}{\bTheta}

\safemath{\matidentity}{\matI}
\safemath{\matone}{\matO}

\safemath{\rnda}{A}
\safemath{\rndb}{B}
\safemath{\rndc}{C}
\safemath{\rndd}{D}
\safemath{\rnde}{E}
\safemath{\rndf}{F}
\safemath{\rndg}{G}
\safemath{\rndh}{H}
\safemath{\rndi}{I}
\safemath{\rndj}{J}
\safemath{\rndk}{K}
\safemath{\rndl}{L}
\safemath{\rndm}{M}
\safemath{\rndn}{N}
\safemath{\rndo}{O}
\safemath{\rndp}{P}
\safemath{\rndq}{Q}
\safemath{\rndr}{R}
\safemath{\rnds}{S}
\safemath{\rndt}{T}
\safemath{\rndu}{U}
\safemath{\rndv}{V}
\safemath{\rndw}{W}
\safemath{\rndx}{X}
\safemath{\rndy}{Y}
\safemath{\rndz}{Z}

\safemath{\rveca}{\bimA}
\safemath{\rvecb}{\bimB}
\safemath{\rvecc}{\bimC}
\safemath{\rvecd}{\bimD}
\safemath{\rvece}{\bimE}
\safemath{\rvecf}{\bimF}
\safemath{\rvecg}{\bimG}
\safemath{\rvech}{\bimH}
\safemath{\rveci}{\bimI}
\safemath{\rvecj}{\bimJ}
\safemath{\rveck}{\bimK}
\safemath{\rvecl}{\bimL}
\safemath{\rvecm}{\bimM}
\safemath{\rvecn}{\bimN}
\safemath{\rveco}{\bomO}
\safemath{\rvecp}{\bimP}
\safemath{\rvecq}{\bimQ}
\safemath{\rvecr}{\bimR}
\safemath{\rvecs}{\bimS}
\safemath{\rvect}{\bimT}
\safemath{\rvecu}{\bimU}
\safemath{\rvecv}{\bimV}
\safemath{\rvecw}{\bimW}
\safemath{\rvecx}{\bimX}
\safemath{\rvecy}{\bimY}
\safemath{\rvecz}{\bimZ}

\safemath{\rvecxi}{\bmxi}
\safemath{\rveclambda}{\bmlambda}
\safemath{\rvecmu}{\bmmu}
\safemath{\rvectheta}{\bmtheta}
\safemath{\rvecphi}{\bmphi}

\safemath{\rmatA}{\bimA}
\safemath{\rmatB}{\bimB}
\safemath{\rmatC}{\bimC}
\safemath{\rmatD}{\bimD}
\safemath{\rmatE}{\bimE}
\safemath{\rmatF}{\bimF}
\safemath{\rmatG}{\bimG}
\safemath{\rmatH}{\bimH}
\safemath{\rmatI}{\bimI}
\safemath{\rmatJ}{\bimJ}
\safemath{\rmatK}{\bimK}
\safemath{\rmatL}{\bimL}
\safemath{\rmatM}{\bimM}
\safemath{\rmatN}{\bimN}
\safemath{\rmatO}{\bimO}
\safemath{\rmatP}{\bimP}
\safemath{\rmatQ}{\bimQ}
\safemath{\rmatR}{\bimR}
\safemath{\rmatS}{\bimS}
\safemath{\rmatT}{\bimT}
\safemath{\rmatU}{\bimU}
\safemath{\rmatV}{\bimV}
\safemath{\rmatW}{\bimW}
\safemath{\rmatX}{\bimX}
\safemath{\rmatY}{\bimY}
\safemath{\rmatZ}{\bimZ}

\safemath{\rmatDelta}{\bimDelta}
\safemath{\rmatLambda}{\bimLambda}
\safemath{\rmatPhi}{\bimPhi}
\safemath{\rmatSigma}{\bimSigma}
\safemath{\rmatOmega}{\bimOmega}
\safemath{\rmatTheta}{\bimTheta}

\usepackage{amssymb}
\usepackage{amsfonts}
\usepackage{mathrsfs}
\usepackage{xspace}
\usepackage{bm}
\usepackage{fancyref}
\usepackage{textcomp}
\usepackage{graphicx}
\usepackage{multirow}
\usepackage{stmaryrd}

\newenvironment{textbmatrix}{	\setlength{\arraycolsep}{2.5pt}%
								\big[\begin{matrix}}{\end{matrix}\big]%
								\raisebox{0.08ex}{\vphantom{M}}}

\def\be{\begin{equation}}
\def\ee{\end{equation}}
\def\een{\nonumber \end{equation}}
\def\mat{\begin{bmatrix}}
\def\emat{\end{bmatrix}}
\def\btm{\begin{textbmatrix}}
\def\etm{\end{textbmatrix}}

\def\ba#1\ea{\begin{align}#1\end{align}}
\def\bas#1\eas{\begin{align*}#1\end{align*}}
\def\bs#1\es{\begin{split}#1\end{split}} 
\def\bg#1\eg{\begin{gather}#1\end{gather}}
\def\bml#1\eml{\begin{multline}#1\end{multline}}
\def\bi#1\ei{\begin{itemize}#1\end{itemize}}

\newcommand{\lefto}{\mathopen{}\left}

\DeclareMathOperator*{\argmin}{arg\;min}		

\newcommand{\vecnorm}[1]{\lefto\lVert#1\right\rVert}		

\safemath{\dirac}{\delta}					
\safemath{\krond}{\dirac}					

\safemath{\upto}{\uparrow}
\safemath{\downto}{\downarrow}
\safemath{\iu}{j}							
\safemath{\ev}{\lambda}						
\safemath{\hilseqspace}{l^{2}}				
\newcommand{\banachfunspace}[1]{\setL^{#1}}	
\safemath{\hilfunspace}{\banachfunspace{2}}	

\safemath{\SNR}{\text{\sc snr}} 				
\safemath{\No}{N_0}							
\safemath{\Es}{E_s}							
\safemath{\Eb}{E_b}							
\safemath{\EbNo}{\frac{\Eb}{\No}}
\safemath{\EsNo}{\frac{\Es}{\No}}

\DeclareMathOperator{\CHop}{\ensuremath{\opH}} 
\safemath{\tvir}{\rndh_{\CHop}}				
\safemath{\tvtf}{\rndl_{\CHop}}				
\safemath{\spf}{\rnds_{\CHop}}				
\safemath{\bff}{H_{\CHop}}					

\safemath{\ircf}{r_{h}}						
\safemath{\tftvcf}{r_{s}}					
\safemath{\tfcf}{r_{l}}						
\safemath{\bfcf}{r_{H}}						

\safemath{\tcorr}{c_h}						
\safemath{\scf}{c_{s}}						
\safemath{\tfcorr}{c_{l}}					
\safemath{\fcorr}{c_{H}}						

\safemath{\mi}{I}							
\safemath{\capacity}{C}						

\safemath{\normal}{\mathcal{N}}			
\safemath{\jpg}{\mathcal{CN}}			
\safemath{\mchain}{\leftrightarrow}		

\safemath{\dB}{\,\mathrm{dB}}
\safemath{\dBm}{\,\mathrm{dBm}}
\safemath{\Hz}{\,\mathrm{Hz}}
\safemath{\kHz}{\,\mathrm{kHz}}
\safemath{\MHz}{\,\mathrm{MHz}}
\safemath{\GHz}{\,\mathrm{GHz}}
\safemath{\s}{\,\mathrm{s}}
\safemath{\ms}{\,\mathrm{ms}}
\safemath{\mus}{\,\mathrm{\text{\textmu}s}}
\safemath{\ns}{\,\mathrm{ns}}
\safemath{\ps}{\,\mathrm{ps}}
\safemath{\meter}{\,\mathrm{m}}
\safemath{\mm}{\,\mathrm{mm}}
\safemath{\cm}{\,\mathrm{cm}}
\safemath{\m}{\,\mathrm{m}}
\safemath{\W}{\,\mathrm{W}}
\safemath{\mW}{\, \mathrm{mW}}
\safemath{\J}{\,\mathrm{J}}
\safemath{\K}{\,\mathrm{K}}
\safemath{\bit}{\,\mathrm{bit}}
\safemath{\nat}{\,\mathrm{nat}}

\safemath{\define}{\triangleq}			

\safemath{\equivalent}{\sim}
\safemath{\distas}{\sim}					
\safemath{\sdiff}{\Delta}				

\safemath{\reals}{\mathbb{R}}
\safemath{\positivereals}{\reals_{+}}
\safemath{\integers}{\mathbb{Z}}
\safemath{\posint}{\integers_{+}}
\safemath{\naturals}{\mathbb{N}}
\safemath{\posnaturals}{\naturals_{+}}
\safemath{\complexset}{\mathbb{C}}
\safemath{\rationals}{\mathbb{Q}}

\newcommand*{\fancyrefapplabelprefix}{app}		
\newcommand*{\fancyrefthmlabelprefix}{thm}		
\newcommand*{\fancyreflemlabelprefix}{lem}		
\newcommand*{\fancyrefcorlabelprefix}{cor}		
\newcommand*{\fancyrefdeflabelprefix}{def}		
\newcommand*{\fancyrefalglabelprefix}{alg}		
\newcommand*{\fancyrefproplabelprefix}{prop}		
\newcommand*{\fancyrefexmpllabelprefix}{exmpl}
\newcommand*{\fancyreftbllabelprefix}{tbl}
\frefformat{vario}{\fancyrefseclabelprefix}{Section~#1}
\frefformat{vario}{\fancyrefthmlabelprefix}{Theorem~#1}
\frefformat{vario}{\fancyreflemlabelprefix}{Lemma~#1}
\frefformat{vario}{\fancyrefcorlabelprefix}{Corrolary~#1}
\frefformat{vario}{\fancyrefdeflabelprefix}{Definition~#1}
\frefformat{vario}{\fancyrefalglabelprefix}{Algorithm~#1}
\frefformat{vario}{\fancyreffiglabelprefix}{Figure~#1}
\frefformat{vario}{\fancyrefapplabelprefix}{Appendix~#1} 
\frefformat{vario}{\fancyrefeqlabelprefix}{(#1)}
\frefformat{vario}{\fancyrefproplabelprefix}{Proposition~#1}
\frefformat{vario}{\fancyrefexmpllabelprefix}{Example~#1}
\frefformat{vario}{\fancyreftbllabelprefix}{Table~#1}

\safemath{\dictab}{[\,\dicta\,\,\dictb\,]}

\safemath{\ysig}{\bmy}
\safemath{\ysighat}{\hat{\ysig}}
\safemath{\ysigdim}{M}
\safemath{\xsig}{\bmx}
\safemath{\xsigdim}{N}
\safemath{\nx}{n_x}
\safemath{\zsig}{\bmz}
\safemath{\zsigdim}{\ysigdim}
\safemath{\rsig}{\bmr}
\safemath{\Adict}{\bA}
\safemath{\Adicttilde}{\widetilde{\Adict}}
\safemath{\Adictdim}{\outputdim\times\xsigdim}
\safemath{\avec}{\bma}
\safemath{\avectilde}{\tilde{\avec}}
\safemath{\Bdict}{\bB}
\safemath{\Bdicttilde}{\widetilde{\Bdict}}
\safemath{\Cdict}{\bC}
\safemath{\cvec}{\bmc}
\safemath{\Ddict}{\bD}
\safemath{\Ddictdim}{\ysigdim\times\xsigdim}
\safemath{\dvec}{\bmd}
\safemath{\Ddicttilde}{\widetilde{\bD}}
\safemath{\Bonb}{\bB}
\safemath{\bvec}{\bmb}
\safemath{\Bonbdim}{\ysigdim\times\ysigdim}
\safemath{\noise}{\bmn}
\safemath{\noisedim}{\ysigim}
\safemath{\err}{\bme}
\safemath{\errdim}{\ysigdim}
\safemath{\errset}{\setE}
\safemath{\nerr}{n_e}
\safemath{\delop}{\bP_\errset}
\safemath{\delopc}{\bP_{{\errset}^c}}

\safemath{\cplxi}{\imath}
\safemath{\cplxj}{\jmath}

\safemath{\dict}{\matD}
\safemath{\inputdim}{N}		
\safemath{\outputdim}{M}		
\safemath{\sparsity}{S}	
\safemath{\inputdimA}{{N_a}}	
\safemath{\inputdimB}{{N_b}}	
\safemath{\elemA}{{n_a}}	
\safemath{\elemB}{{n_b}}	
\safemath{\resA}{\matR_a}	
\safemath{\resB}{\matR_b}	
\safemath{\subD}{\matS} 
\safemath{\subA}{\matS_a} 
\safemath{\subB}{\matS_b} 
\safemath{\dicta}{\matA} 	
\safemath{\dictb}{\matB} 	
\safemath{\hollowS}{H}
\safemath{\hollowA}{H_a}
\safemath{\hollowB}{H_b}
\safemath{\cross}{Z}
\safemath{\coh}{\mu_d}			
\safemath{\coha}{\mu_a}			
\safemath{\cohb}{\mu_b}			
\safemath{\mubs}{\nu}	
\safemath{\cohm}{\mu_m} 
\safemath{\dictset}{\setD}	
\safemath{\dictsetp}{\dictset(\coh,\coha,\cohb)}	
\safemath{\dictsetgen}{\dictset_\text{gen}}
\safemath{\dictsetgenp}{\dictsetgen(\coh)}
\safemath{\dictsetonb}{\dictset_\text{onb}}
\safemath{\dictsetonbp}{\dictsetonb(\coh)}

\safemath{\leftside}{U}
\safemath{\rightsideA}{R_a}
\safemath{\rightsideB}{R_b}

\safemath{\indexS}{\setI_S} 

\safemath{\na}{n_a}			
\safemath{\nb}{n_b}			
\safemath{\coeffa}{p_i}	
\safemath{\coeffb}{q_j}	
\safemath{\seta}{\setP}		
\safemath{\setb}{\setQ}     
\safemath{\setw}{\setW}	
\safemath{\setz}{\setZ}	
\safemath{\cola}{\veca}		
\safemath{\colb}{\vecb}		
\safemath{\cold}{\vecd}		
\safemath{\inputvec}{\vecx} 	
\safemath{\error}{\vece}	
\safemath{\noiseout}{\vecz} 	
\safemath{\inputvecel}{x}
\safemath{\inputveca}{\vecx_a}
\safemath{\inputvecb}{\vecx_b}
\safemath{\outputvec}{\vecy}	
\safemath{\lambdamin}{\lambda_{\mathrm{min}}}

\safemath{\elltwo}{\ell_2}
\safemath{\ellone}{\ell_1}
\safemath{\ellzero}{\ell_0}
\safemath{\ellinf}{\ell_\infty}
\safemath{\licard}{Z(\coh,\coha,\cohb)}
\safemath{\xsol}{\hat{x}}
\safemath{\xbord}{x_b}		
\safemath{\xstat}{x_s}		
\safemath{\xstatLone}{\tilde{x}_s}
\safemath{\order}{\mathcal{O}} 
\safemath{\scales}{\Theta} 
\safemath{\ones}{\mathbf{1}} 
\safemath{\zeroes}{\mathbf{0}} 
\safemath{\thlone}{\kappa(\coh,\cohb)} 
\safemath{\constoneA}{\delta} 
\safemath{\constoneB}{\epsilon} 
\safemath{\nlarge}{L}				   
\safemath{\sumlarge}{S_\nlarge}
\safemath{\maxlarger}{P_\nlarge}	   
\safemath{\Pzero}{\textrm{P0}}	
\safemath{\Pone}{\textrm{P1}}
\safemath{\vecfir}{\vecw}			 
\safemath{\vecsec}{\vecz}
\safemath{\elvecfir}{w}              
\safemath{\elvecsec}{z}				 
\safemath{\nlargefir}{n}
\safemath{\normout}{\gamma}
\safemath{\auxfun}{h}
\safemath{\supp}{\textrm{supp}}

\safemath{\indexa}{\ell}
\safemath{\indexb}{r}
\safemath{\indexc}{i}
\safemath{\indexd}{j}

\safemath{\project}{P}

\IEEEoverridecommandlockouts

\begin{document}
\begin{spacing}{0.9}
\title{Decentralized Coordinate-Descent Data Detection \\ and Precoding for Massive MU-MIMO}

\bstctlcite{IEEEexample:BSTcontrol}
\setlength{\textfloatsep}{3pt}%

\author{\IEEEauthorblockN{Kaipeng Li$^{1}$, Oscar Casta\~neda$^{2}$, Charles Jeon$^{3}$,  Joseph R. Cavallaro$^{1}$, and Christoph Studer$^{2}$\thanks{The work was supported in part by Xilinx, Inc.~and by the US NSF under grants ECCS-1408370, CNS-1717218, CNS-1827940, ECCS-1408006, CCF-1535897, CCF-1652065, and CNS-1717559. We acknowledge the hardware support of the DGX-1 multi-GPU systems at the Nvidia Data Center.}}
\IEEEauthorblockA{$^{1}$Department of Electrical and Computer Engineering, Rice University, Houston, TX\\
    $^{2}$School of Electrical and Computer Engineering, Cornell University, Ithaca, NY\\
    $^{3}$Intel Labs, Hillsboro, OR\\
    \vspace{-0.8cm}    }}

\maketitle
\begin{abstract}
Massive multiuser (MU) multiple-input multiple-output (MIMO) promises significant improvements in spectral efficiency compared to small-scale MIMO. Typical massive MU-MIMO base-station (BS) designs rely on centralized linear data detectors and precoders which entail excessively high complexity, interconnect data rates, and chip input/output (I/O) bandwidth when executed on a single computing fabric. To resolve these complexity and bandwidth bottlenecks, we propose new decentralized algorithms for data detection and precoding that use coordinate descent. Our methods parallelize computations across multiple computing fabrics, while minimizing interconnect and I/O bandwidth. The proposed decentralized algorithms achieve near-optimal error-rate performance and multi-Gbps throughput at sub-1\,ms latency when implemented on a multi-GPU cluster with half-precision floating-point arithmetic.
\end{abstract}

\section{Introduction}
\label{sec:intro}
Massive multi-user (MU) multiple-input multiple-output (MIMO) will be a key technology of next-generation wireless systems~\cite{LETM2014,ABCHLAZ2014}. Equipped with hundreds of antennas, a massive MU-MIMO base-station (BS) simultaneously serves tens of user equipments (UEs) in the same time-frequency resource; this yields significant spectral efficiency and power efficiency improvements compared to that of conventional, small-scale MIMO systems. 
As discussed in~\cite{li2017decentralized}, centralized baseband processing in massive MU-MIMO results in excessively high complexity  as well as interconnect and chip input/output (I/O) bandwidth between the baseband processors and the antenna array. 
For example, 
the raw baseband data rates  for a massive MU-MIMO BS operating at 80\,MHz bandwidth with 256 active antenna elements and 12-bit digital-to-analog converters (DACs) approach 1\,Tbps, exceeding the capabilities of existing interconnect standards, such as the Common Public Radio Interface (CPRI)~\cite{cpri}. Furthermore, a single centralized computing fabric, such as a field programmable gate array (FPGA) or a graphics processing unit (GPU), has limited chip I/O bandwidth, computing, and storage resources to realize real-time baseband processing at such high rates.

\subsection{Decentralized Baseband Processing}
\label{sec:dbp}
To resolve the interconnect and I/O bandwidth bottlenecks and reduce complexity and memory requirements per computing fabric, 
 references~\cite{li2017decentralized,EBDistribute2018,EBasilomar16,MAMMOET} have proposed \emph{decentralized baseband processing} (DBP) for massive MU-MIMO systems. The key idea behind DBP is to divide the antenna array into separate antenna clusters, each associated with dedicated RF circuitry and baseband processors. At each cluster, local baseband processing, such as (de-)modulation, channel estimation, data detection, and  precoding is performed. To achieve near-optimal spectral efficiency, algorithms that rely on consensus-sharing among antenna clusters~\cite{LCSGCS2016,LSCCGS2016} or one-shot  feedforward architectures~\cite{LJCS2018,jeon2017achievable,CJTSP2018} have been developed.  
Existing algorithms for DBP, however, require costly matrix-matrix multiplications and matrix inversion operations~\cite{jeon2017achievable,CJTSP2018} at each cluster, which causes high complexity in scenarios with  short coherence times that require channel matrix preprocessing at high rates.

\subsection{Contributions}
\label{sec:contrib}
\begin{figure}[t]
\centering
\subfigure[Uplink equalization]{\includegraphics[width=0.24\textwidth]{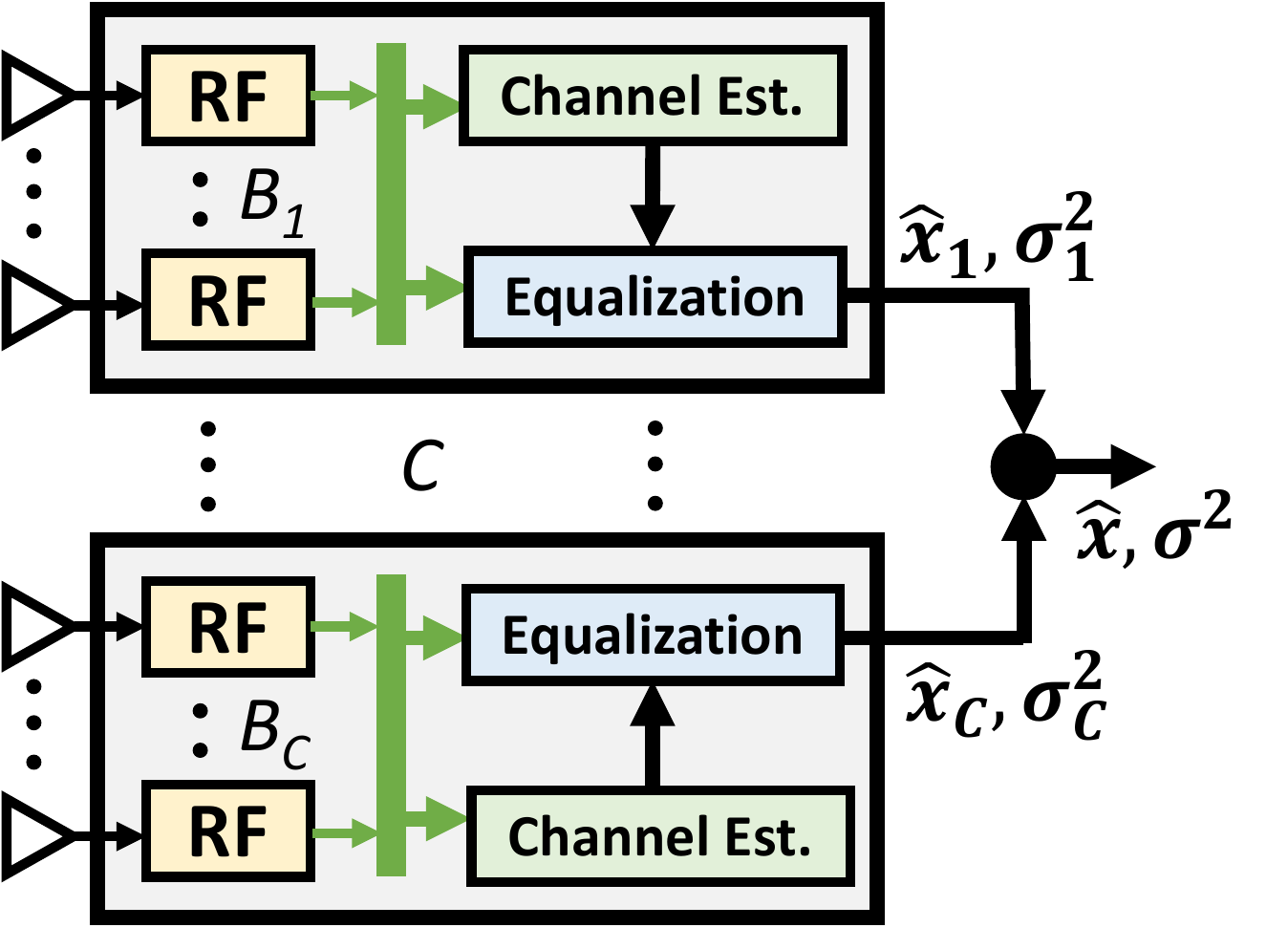}\label{fig:1a}}
\subfigure[Downlink precoding]{\includegraphics[width=0.225\textwidth]{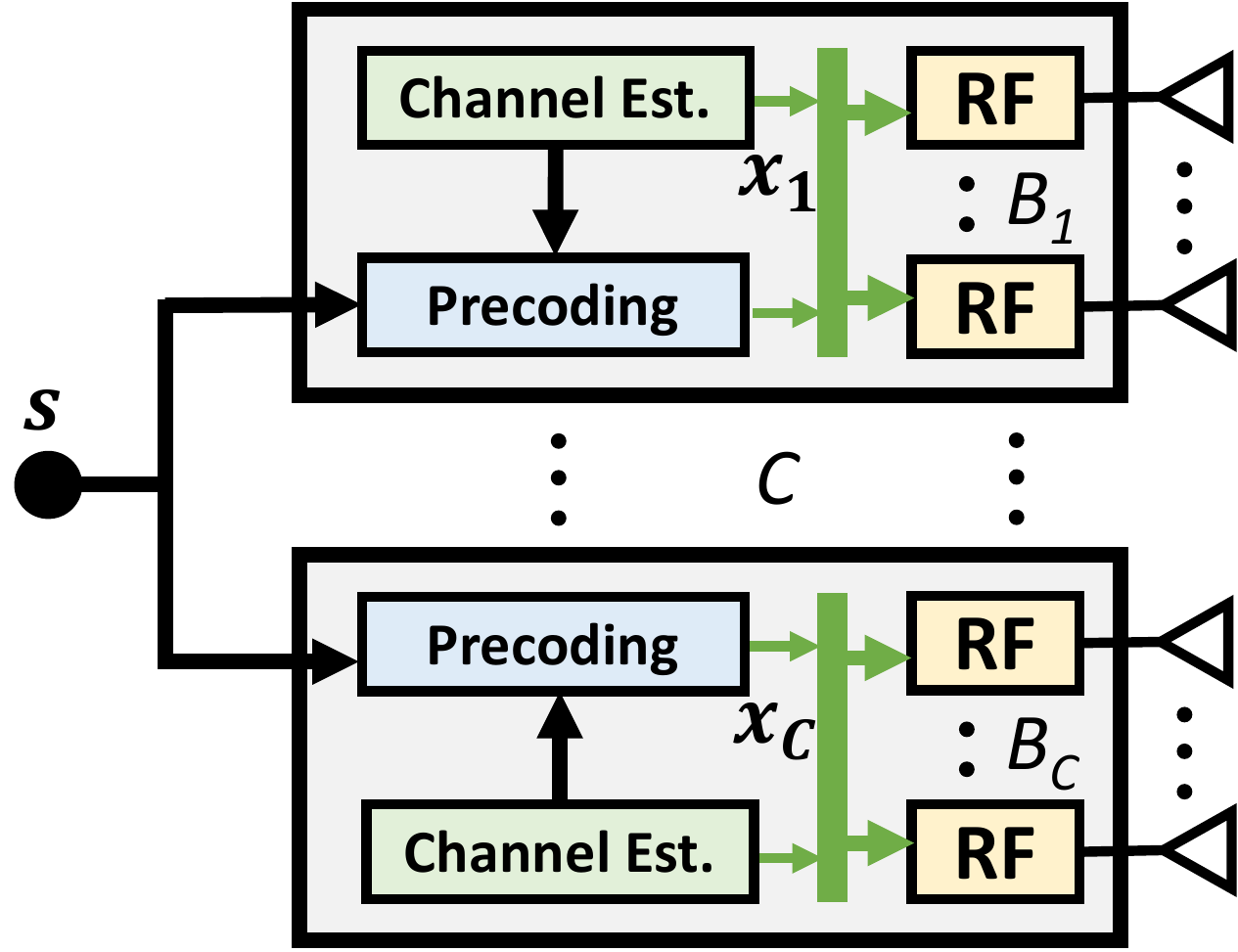}\label{fig:1b}}

\caption{Fully-decentralized feedforward architectures for massive MU-MIMO.}
\label{fig:fdarch}
\end{figure}

We propose new coordinate descent~(CD)-based data detection and  precoding algorithms that leverage fully-decentralized \emph{feedforward} architectures as illustrated in Fig~\ref{fig:fdarch}. 
In the uplink (UEs transmit to BS), we perform linear minimum mean-square error (L-MMSE) equalization using CD at each cluster, and fuse the local estimates from all clusters to form a global detection result. 
In the downlink (BS transmits to UEs), we perform zero-forcing (ZF) precoding using CD at each cluster and allocate the power per local clusters under a global power constraint. 
Our decentralized CD-based data detector and precoder avoid computation of the Gram matrix and matrix inversion, which yields significant complexity savings while maintaining near-optimal error-rate performance. To showcase the efficacy of our methods in practice, we implement our algorithms on a multi-GPU system using both single-precision (32-bit) and half-precision (16-bit) floating-point formats. Our implementations outperform existing centralized and decentralized methods, and show that multi-Gbps throughput at sub-1ms latency can be achieved for realistic massive MU-MIMO systems.

\begin{algorithm}[t]
\caption{Decentralized L-MMSE Detection using CD
 \label{alg:dcd_det}}
\small
\begin{algorithmic}
\STATE \textbf{Input:} $\mathbf{H}_c$, $\mathbf{y}_c$, $c=1,\ldots,C$ and $N_0, E_x$
\STATE \textbf{\em Preprocessing (decentralized):}

\STATE $m_{u,c} = (\vecnorm{\bmh_{u,c}}_2^2+\frac{N_0}{E_x})^{-1}, u=1,\ldots,U, c=1,\ldots,C$
\STATE $n_{u,c} = m_{u,c}\vecnorm{\bmh_{u,c}}_2^2, u=1,\ldots,U, c=1,\ldots,C$

\STATE \textbf{\em CD iterations (decentralized):}
\STATE \textbf{Init:} $\bmr_c=\bmy_c, \mathbf{x}_c^{(0)}=\mathbf{0}$

\FOR{$t = 1,\ldots, T_\text{max}$}
\FOR{$u = 1,\ldots, U$}
\STATE $x_{u,c}^{(t)}=m_{u,c}\bmh_{u,c}^H\bmr_c+n_{u,c}x_{u,c}^{(t-1)}$
\STATE $\delta x_{u,c}^{(t)} = x_{u,c}^{(t)} - x_{u,c}^{(t-1)}$
\STATE $\bmr_c = \bmr_c - \bmh_{u,c}\delta x_{u,c}^{(t)}$
\ENDFOR
\ENDFOR
\STATE \textbf{\em Data fusion and averaging (centralized):}
\STATE \textbf{Output:} $\hat\bmx = \Sigma_{c=1}^{C}{\lambda_c\hat\bmx_c^{(T_\text{max})}}$
\end{algorithmic}

\end{algorithm}

\section{System Model, Architecture, and Algorithms}
\label{sec:arch}
\subsection{Uplink System Model and Architecture}
\label{sec:ulmodel}
In the uplink, $U$ single-antenna UEs transmit the vector $\mathbf{x}^\text{ul}\in\setO^U$ ($\setO$ is the constellation set) to the BS with $B$ antennas, where $B\geq U$. The received signal $\mathbf{y}^\text{ul}\in\mathbb{C}^B$ at the BS can be modeled using the well-known baseband input-output relation $\bmy^\text{ul}=\bH^\text{ul}\bmx^\text{ul}+\bmn^\text{ul}$, where $\bH^\text{ul}\in\mathbb{C}^{B\times U}$ is the MIMO channel matrix, and $\bmn^\text{ul}\in \mathbb{C}^{B}$ models noise as i.i.d.\ complex circularly-symmetric Gaussian random vector with variance $N_0$ per entry. 

Recently, fully-decentralized uplink architectures have been proposed in~\cite{CJTSP2018,jeon2017achievable}. As shown in Fig.~\ref{fig:1a}, the $B$ BS-antennas are partitioned into $C$ antenna clusters, where the $c^\text{th}$ cluster consists of $B_c$ antennas with $B=\sum^C_{c=1}{B_c}$. Each cluster receives its own signals $\bmy_c^\text{ul}\in\mathbb{C}^{B_c}$ and the corresponding input-output relation for each cluster is 
\begin{equation*}
\bmy_c^\text{ul}=\bH_c^\text{ul}\bmx^\text{ul}+\bmn_c^\text{ul}, \quad c=1,2,\ldots,C.
\end{equation*}
Here, $\bH_c^\text{ul}\in \mathbb{C}^{B_c\times U}$ (a sub-matrix of $\bH^\text{ul}$) is the local channel matrix and $\bmn_c^{ul}\in \mathbb{C}^{B_c}$ is the local noise vector at cluster $c$.

For the fully-decentralized feedforward uplink architecture shown in Fig.~\ref{fig:1a}, each BS cluster estimates and stores its own matrix $\bH_c^\text{ul}$, and performs local data detection based only on its local receive vector $\bmy_c^\text{ul}$ and channel matrix $\bH_c^\text{ul}$ to form a local estimate. 
All clusters then send their local estimates $\hat{\bmx}_c^\text{ul}$ in a feedforward manner to a centralized processor that forms the global estimate $\hat{\bmx}^\text{ul}$. See Sec.~\ref{sec:detection} for more details.

\subsection{Downlink System Model and Architecture}
\label{sec:dlmodel}
In the downlink, the BS precodes the vector $\bms\in\setO^U$ to form a beamforming vector $\bmx^\text{dl}\in\mathbb{C}^B$ that is transmitted to $U$ UEs. The UE vector $\bmy^\text{dl}\in\mathbb{C}^U$ that contains the receive signal for each UE is given by $\bmy^\text{dl}=\bH^\text{dl}\bmx^\text{dl}+\bmn^\text{dl}$, where $\bH^\text{dl}\in\mathbb{C}^{U\times B}$ and $\bmn^\text{dl}\in \mathbb{C}^{U}$ are the downlink channel and noise, respectively. 

In the fully decentralized downlink architecture put forward in~\cite{LJCS2018} and shown in Fig.~\ref{fig:1b}, there are $B_c$ BS antennas at each of the $C$ clusters. The local input-output relation is
 \begin{equation*}
\bmy_c^\text{dl}=\bH_c^\text{dl}\bmx_c^\text{dl}+\bmn_c^\text{dl}, \quad c=1,2,\ldots ,C.
\end{equation*}
Here, $\bmx_c^\text{dl}\in\mathbb{C}^{B_c}$ is the local beamforming vector and the local downlink channel can be obtained via reciprocity in the uplink, i.e., $\bH_c^\text{dl}={\bH_c^\text{ul}}^H$. In Sec.~\ref{sec:precoding}, we show a new CD-based precoding algorithm for the fully-decentralized architecture to compute~$\bmx_c^\text{dl}$ in a feedforward manner.

\begin{algorithm}[t]
\caption{Decentralized ZF Precoding using CD\label{alg:dcd_pre}}
\small
\begin{algorithmic}
\STATE \textbf{Input:} $\mathbf{H}_c$, $\mathbf{s}$, $\rho_c$, $c=1,\ldots,C$
\STATE \textbf{\em Data broadcasting (centralized node $\rightarrow$ all clusters)}: $\mathbf{s}$
\STATE \textbf{\em Preprocessing (decentralized):}
\STATE $p_{u,c} = \vecnorm{\bmh_{u,c}}_2^{-1}, u=1,\ldots,U, c=1,\ldots,C$
\STATE $\bar{\bmh}_{u,c} = p_{u,c}\bmh_{u,c}, u=1,\ldots,U, c=1,\ldots,C$
\STATE $\bar{s}_u = p_{u,c}s_u, u=1,\ldots,U, c=1,\ldots,C$

\STATE \textbf{\em CD iterations (decentralized):}
\STATE \textbf{Init:} $\mathbf{x}_c=\mathbf{0}$

\FOR{$t = 1,\ldots, T_\text{max}$}
\FOR{$u = 1,\ldots, U$}
\STATE $\bmx_c = \bmx_c - (\bar{\bmh}_{u,c}\bmx_c-\bar{s}_u)\bar{\bmh}_{u,c}^H$
\ENDFOR
\ENDFOR
\STATE $\bmx_c = {\rho_c\bmx_c}/{\vecnorm{\bmx_c}_2}$ \textbf{\em(decentralized)}
\STATE \textbf{Output:} $\hat\bmx = [\bmx_1; \ldots; \bmx_C]$

\end{algorithmic}

\end{algorithm}

\begin{figure*}[t]
\centering
\subfigure[$C=4$, $B=128$, uplink]{\includegraphics[width=0.24\textwidth]{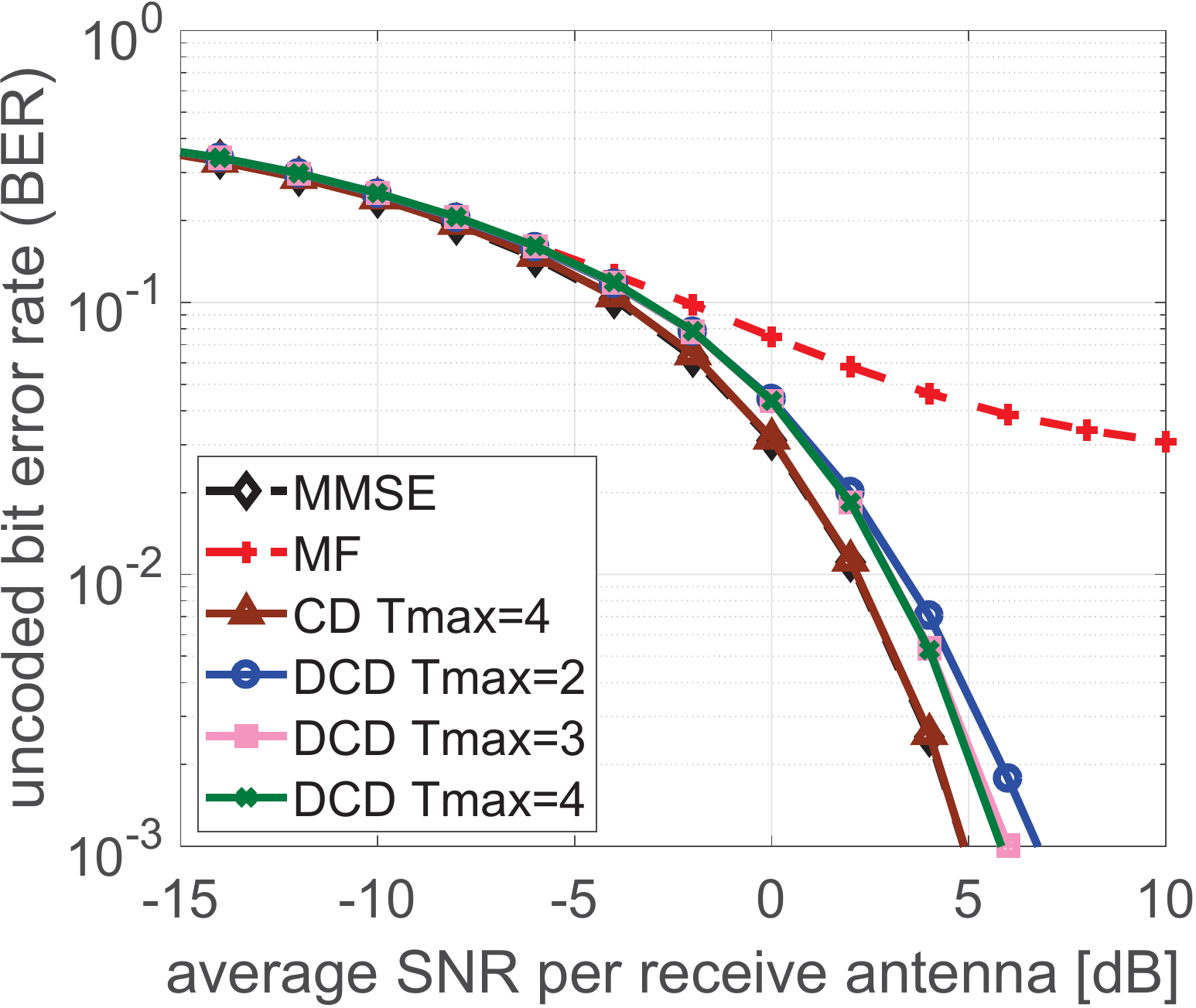}\label{fig:2a}}
\subfigure[$C=8$, $B=256$, uplink]{\includegraphics[width=0.24\textwidth]{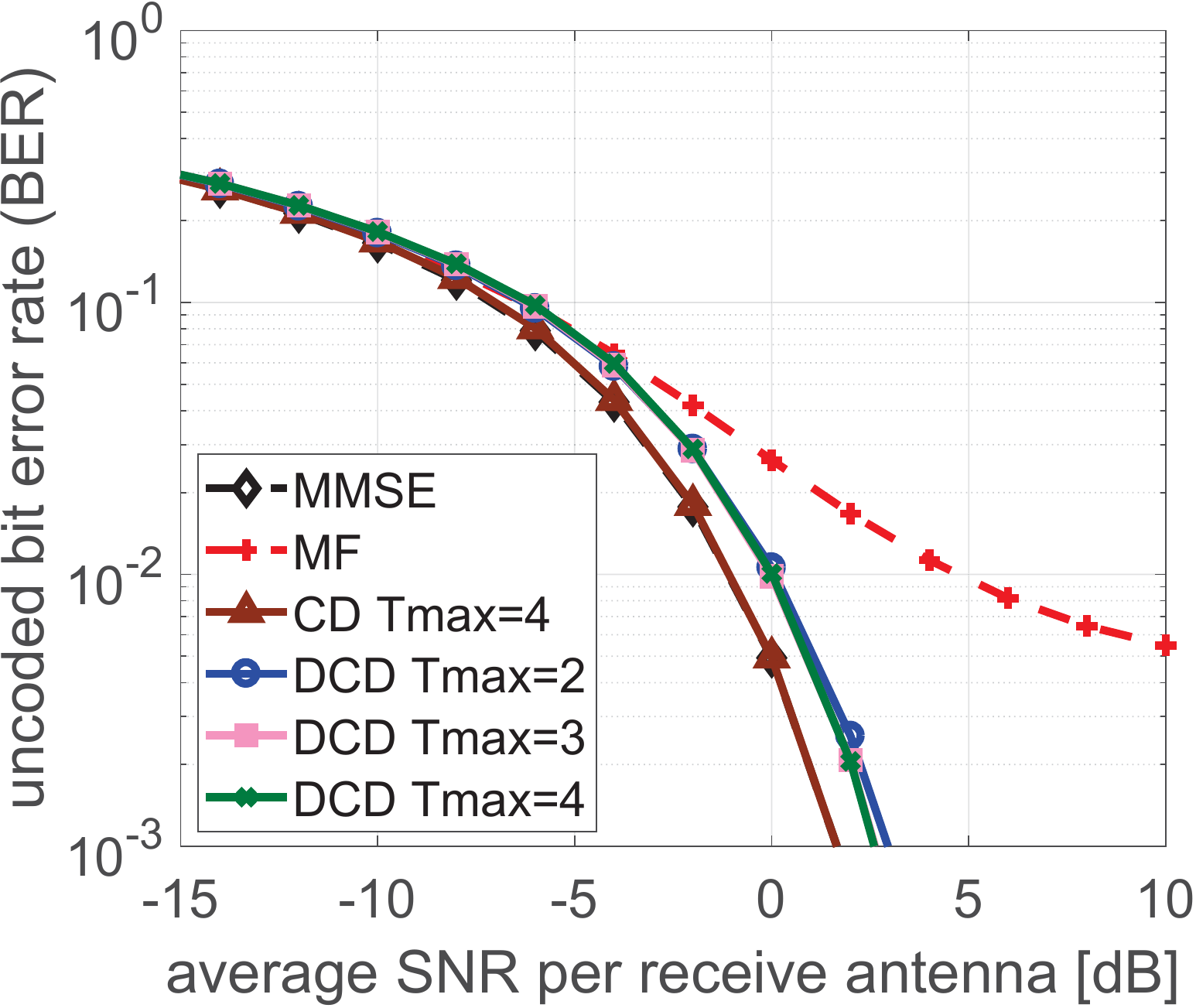}\label{fig:2b}}
\subfigure[$C=4$, $B=128$, downlink]{\includegraphics[width=0.24\textwidth]{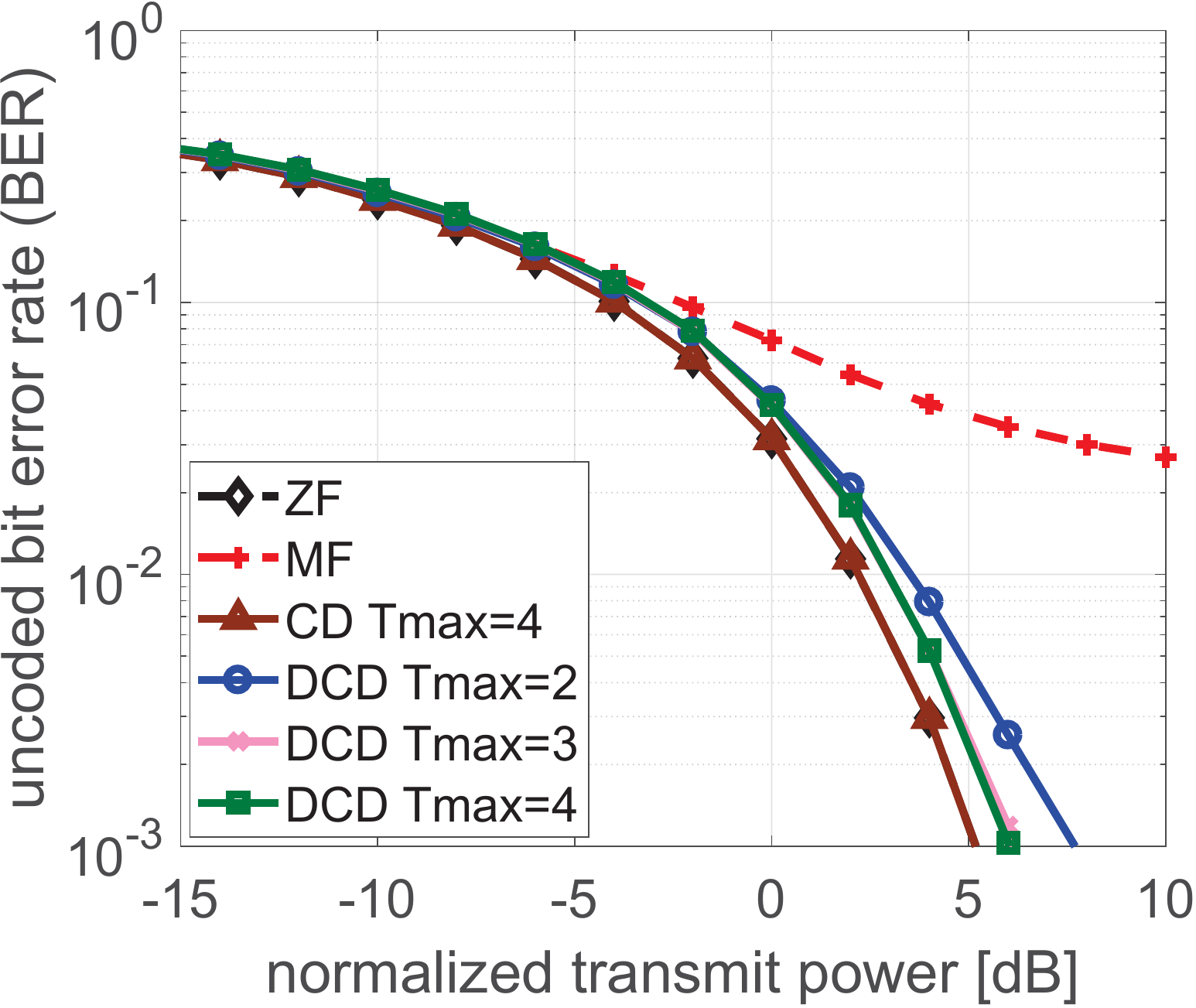}\label{fig:2c}}
\subfigure[$C=8$, $B=256$, downlink]{\includegraphics[width=0.24\textwidth]{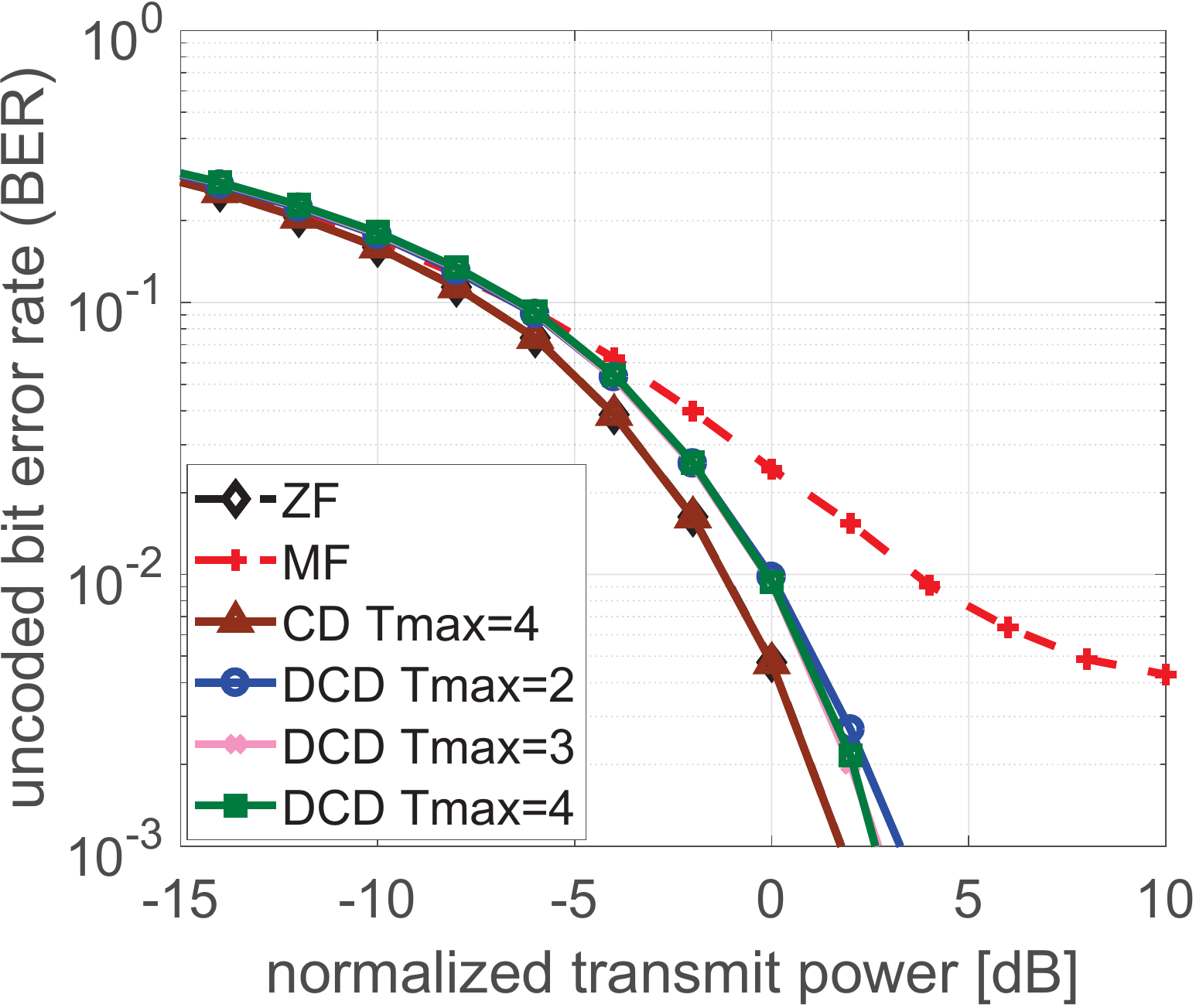}\label{fig:2d}}
\caption{Uncoded bit error rate (BER) performance for $U=8$ users and $B_c=32$ antennas per cluster. Figs. \ref{fig:2a} and \ref{fig:2b}: Uplink data detection performance comparison between decentralized CD, MF, centralized CD and L-MMSE data detectors. Figs. \ref{fig:2c} and \ref{fig:2d}: Downlink precoding performance comparison between decentralized CD, MF, centralized CD and ZF precoders. We scale  the total BS antenna number $B$ by increasing the cluster size $C$.}
\label{fig:berfig}
\vspace{-10pt}
\end{figure*}

\section{Decentralized Uplink Data Detection}
\label{sec:detection}
In the uplink, the BS estimates the transmit signal $\hat{\bmx}^\text{ul}$ based on $\bmy^\text{ul}$ and $\bH^\text{ul}$. In what follows,  we omit the superscript $^\text{ul}$ and we focus on the L-MMSE data detector by solving the following convex optimization problem:
\begin{align}
\hat\bmx = \argmin_{\bmx\in\complexset^U} \, \textstyle  \vecnorm{\bmy-\bH\bmx}_2^2 + \frac{N_0}{E_x}\vecnorm{\bmx}_2^2.
\end{align}
Here, $E_x$ represents the per-user transmit energy. To solve this problem efficiently, reference \cite{WDCS2016} proposed to use coordinate descent (CD), which avoids computation of the Gram matrix and  matrix inversion, and effectively recovers $\hat{\bmx}$ with only a few iterations for massive MU-MIMO systems. 
We propose to decentralize this CD-based detector by leveraging the fully-decentralized feedforward architecture in Fig.~\ref{fig:1a}.
 At each antenna cluster, we calculate a local estimate~$\hat\bmx_c$ given $\bmy_c$ and $\bH_c$ using CD as in~\cite{WDCS2016}. We then fuse the local estimates $\hat\bmx_c$ to form a global estimate $\hat\bmx$ via a weighted sum: $\hat\bmx = \Sigma_{c=1}^{C}{\lambda_c\hat\bmx_c}$, where $\lambda_c = \frac{1}{\sigma_c^2}(\Sigma_{c'=1}^C{\frac{1}{\sigma_{c'}^2}})^{-1}$ is the optimal fusion rule proposed in \cite{jeon2017achievable}; $\sigma_{c}^2$ is the post-equalization noise variance at cluster $c$. The decentralized CD data detector is summarized in Algorithm~\ref{alg:dcd_det}. {As seen from  Algorithm~\ref{alg:dcd_det}, the CD based detector does not require computationally-intensive Gram matrix computations and matrix inversions; instead, the algorithm mostly relies on vector operations, which  significantly reduces the complexity of L-MMSE data detection.}

\section{Decentralized Downlink Precoding}
\label{sec:precoding}

\subsection{Coordinate Descent (CD)-based Precoding}
\label{sec:cdbeam}
In the downlink, the BS precodes the transmit signal $\bms$ using the matrix $\bH^\text{dl}$ to generate a beamforming vector $\bmx^\text{dl}$. In what follows, we omit the superscript $^\text{dl}$ and focus on ZF precoding which minimizes MU interference. ZF precoding can be formulated as the following convex optimization problem: 
\begin{align}
\label{eq:zf_precoding}
\hat\bmx = \argmin_{\bmx\in\complexset^B} \textstyle \frac{1}{2}\vecnorm{\bmx}_2^2 \quad \text{subject to} \,\, \bms=\bH\bmx,
\end{align}
which has a closed form solution $\hat\bmx=\bH^H(\bH\bH^H)^{-1}\bms$. We next derive a CD-based ZF precoder that avoids the need of matrix multiplications and inversions. 
To simplify the derivation, we normalize $\bH$ so that the $u^\text{th}$ row of $\bar{\bH}$ is $\bar{\bmh}_u^H = \bmh_u^H/\vecnorm{\bmh_u}_2$. We also scale $\bms$ accordingly so that the $u^\text{th}$ entry of $\bar{\bms}$ is $\bar{s}_u=s_u/\vecnorm{\bmh_u}_2$. We note that the scaled beamforming constraint $\bar{\bms} = \bar{\bH}\bmx$ can be used in \eqref{eq:zf_precoding} without altering the solution.  
We first reformulate~\eqref{eq:zf_precoding} to its Lagrangian dual
\begin{align}
\label{eq:dual}
\hat\bmz = \argmin_{\bmz\in\complexset^U} f(\bmz):=\textstyle \frac{1}{2}\vecnorm{\bar{\bH}^H\bmz}^2_2-\bar{\bms}^H\bmz
\end{align}
for which the primal solution $\hat\bmx$ is given by $\hat\bmx = \bar{\bH}^H\hat\bmz$. We solve \eqref{eq:dual} using CD by updating $\bmz$ iteratively. Specifically, we update~$\bmz$ across each of its $U$ coordinates in the opposite direction to the gradient component of that coordinate, respectively, in round-robin fashion. For coordinate $u$, we update $\bmz$ as
\begin{align}
\bmz \leftarrow \bmz-\nabla f(\bmz)_u\bme_u =\bmz - (\bar{\bmh}_u^H\bar{\bH}^H\bmz-\bar{s}_u)\bme_u,
\end{align}
where $\bme_u$ is the $u^\text{th}$ unit vector. Given $\hat\bmx = \bar{\bH}^H\hat\bmz$, we have:
\begin{align}
\label{eq:updating}
\bmx \leftarrow \bmx - (\bar{\bmh}_u^H\bar{\bH}^H\bmz-\bar{s}_u)\bar{\bmh}_u = \bmx - (\bar{\bmh}_u^H\bmx-\bar{s}_u)\bar{\bmh}_u,
\end{align}
which is the update formula for the $u^\text{th}$ coordinate. We iteratively update $\bmx$ with \eqref{eq:updating} across all $U$ coordinates in a cyclic fashion until convergence.
Once $\bmx$ is computed, we transmit $\bmx\leftarrow{\rho\bmx}/{\vecnorm{\bmx}_2}$ to satisfy the transmit power constraint~$\rho^2$.

\begin{figure}[t]
\centering
\includegraphics[width=0.57\columnwidth]{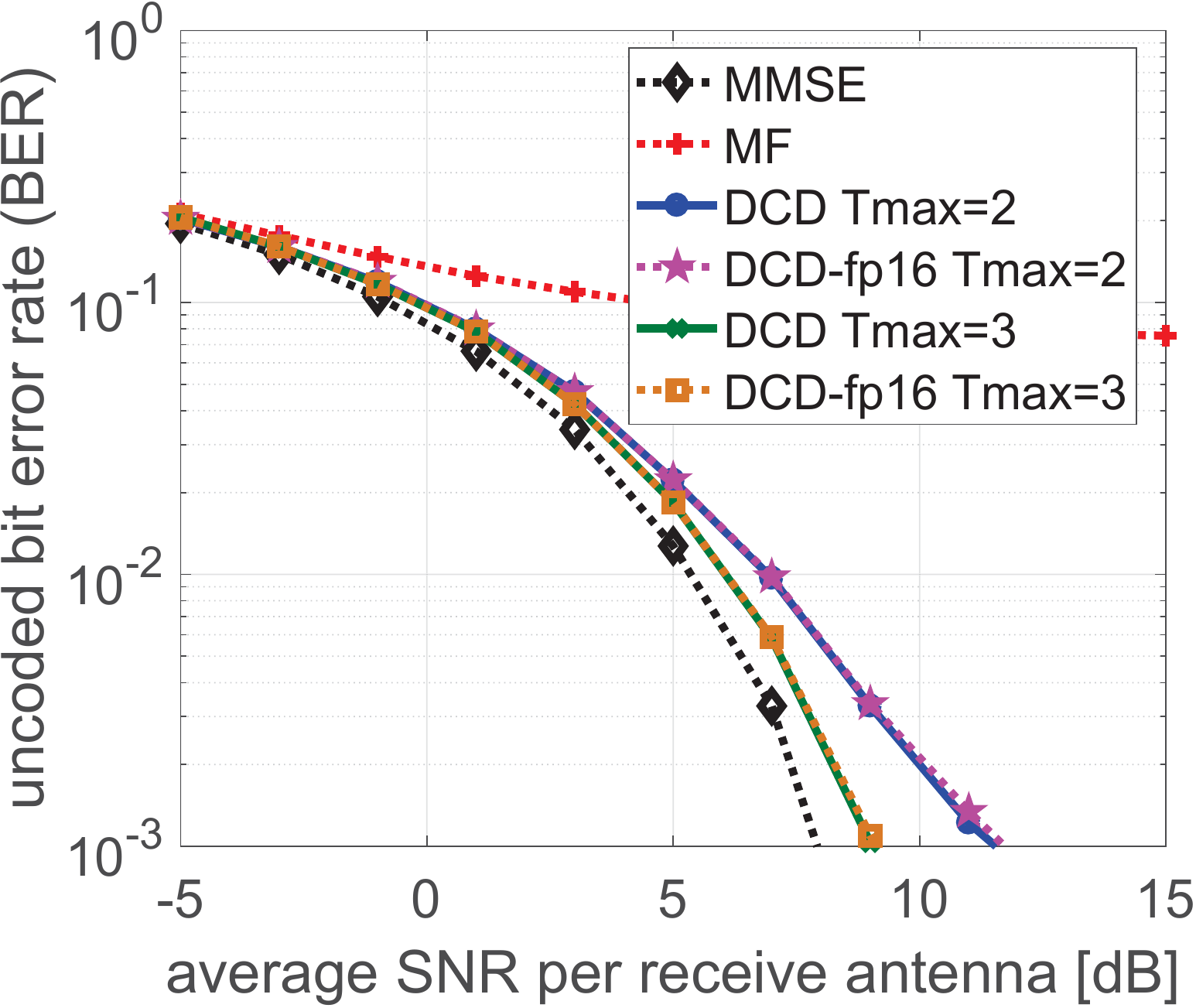}
\vspace{-0.1cm}
\caption{BER of 16-bit vs.\ double precision floating-point uplink detectors.}
\label{fig:fp16}
\end{figure} 

\begin{table*}
\centering
\small
\caption{Latency (L) and Throughput (TP) performance of decentralized CD data detection
and precoding}
\vspace{-0.1cm}
\begin{tabular}{@{}lcccccccc@{}}
\toprule 
 & \multicolumn{4}{c}{Uplink data detection} & \multicolumn{4}{c}{Downlink precoding}\tabularnewline
\midrule 
$U=8$, $B_c=32$ & \multicolumn{2}{c}{$C=4$, $B=128$} & \multicolumn{2}{c}{$C=8$, $B=256$} & \multicolumn{2}{c}{$C=4$, $B=128$} & \multicolumn{2}{c}{$C=8$, $B=256$}\tabularnewline
L [ms] / TP [Gbps] &fp32 & fp16 & fp32 & fp16 & fp32 & fp16 & fp32 & fp16\tabularnewline
\midrule
$T_\text{max}=2$ & 0.465/1.23 & 0.259/2.21 & 0.502/1.14 & 0.295/1.95 & 0.414/1.38 & 0.228/2.52 & 0.440/1.30 & 0.233/2.46\tabularnewline
$T_\text{max}=3$ & 0.540/1.06 & 0.298/1.92 & 0.585/0.98 & 0.331/1.73 & 0.499/1.15 & 0.267/2.15 & 0.525/1.09 & 0.275/2.09\tabularnewline
$T_\text{max}=4$ & 0.617/0.93 & 0.341/1.68 & 0.665/0.86 & 0.370/1.55 & 0.577/0.99 & 0.305/1.88 & 0.601/0.95 & 0.311/1.84\tabularnewline
\bottomrule 
\end{tabular}
\label{tb:datarate}
\vspace{-8pt}
\end{table*}

\begin{table*}
\small
\centering
\caption{Performance comparison between different Massive MU-MIMO data detectors (-D) and precoders (-P)}
\vspace{-0.1cm}
\begin{tabular}{@{}lcccccc@{}}
\toprule 
 & Neumann-D~\cite{BYICASSP13} & OCD-D~\cite{WDCS2016} & FD-LAMA-D~\cite{CJTSP2018} & \textbf{DCD-D} & FD-WF-P~\cite{LJCS2018} & \textbf{DCD-P}\tabularnewline
\midrule 
Fabric & ASIC (fixed-point) & FPGA (fixed-point) & GPU (fp32) & \textbf{GPU (fp16)} & GPU (fp32) & \textbf{GPU (fp16)}\tabularnewline
Archtecture & Centralized & Centralized & Decentralized & \textbf{Decentralized} & Decentralized & \textbf{Decentralized}\tabularnewline
Antenna & $128\times 8$ &  $128\times 8$ &  $128\times 16$ &  $\mathbf{128\times 8}$ &  $128\times 16$ &  $\mathbf{128\times 8}$\tabularnewline
Modulation & 64-QAM & 64-QAM & 16-QAM & \textbf{16-QAM} & 64-QAM & \textbf{16-QAM}\tabularnewline
Iteration & 3 & 3 & 3 & \textbf{3} & N/A & \textbf{3}\tabularnewline
Throughput (TP) & 3.8 Gbps & 0.38 Gbps & 1.34 Gbps & \textbf{1.92 Gbps} & 1.91 Gbps & \textbf{2.15 Gbps}\tabularnewline
TP/UE\,@\,16-QAM  & 317 Mbps & 31 Mbps & 84 Mbps & \textbf{240 Mbps} & 80 Mbps & \textbf{269 Mbps}\tabularnewline
\bottomrule 
\end{tabular}
\label{tb:perfcompare}
\vspace{-10pt}
\end{table*}

\subsection{Decentralized CD-based Precoding}
\label{sec:dcdbeam}
The proposed decentralized CD precoder is suitable for the fully-decentralized feedforward architecture depicted in Fig.~\ref{fig:1b}. 
Given a transmit power constraint $\rho^2$, we first broadcast the transmit signal $\bms$ to each cluster, and solve the local ZF precoding problem using CD as described above. We then scale the beamforming vector $\bmx_c$ at each cluster to the local power constraint $\rho_c^2={\rho^2}/{C}$ so that $\bmx_c \leftarrow {\rho_c\bmx_c}/{\vecnorm{\bmx_c}_2}$.  The decentralized CD precoder is summarized in Algorithm~\ref{alg:dcd_pre}. \revision{As for data detection, Gram matrix computation and matrix inversion are avoided, which reduces complexity.}

\section{Simulation Results}
\label{sec:ber}
We now show uncoded bit error-rate (BER) results of the proposed decentralized CD-based data detection and precoding algorithms for a Rayleigh fading massive MU-MIMO system with 16-QAM. We  fix the number of local BS antennas $B_c=32$, and the numbers of UEs $U=8$. Figs.~\ref{fig:2a} and \ref{fig:2b} show the results of data detection for a total number of BS antennas $B=\{128, 256\}$ divided into $C=\{4,8\}$ clusters, respectively; Figs.~\ref{fig:2c} and \ref{fig:2d} show the results for precoding. We see that for data detection and precoding, the proposed decentralized CD-based methods effectively approach the performance of centralized methods with  $T_\text{max}=3$ or $4$ iterations and with negligible performance loss. \revision{For example, our methods suffer only a 1\,dB SNR loss at $10^{-3}$ BER,} while outperforming the fully-decentralized matched filter (MF) by a significant margin. 

To illustrate the effect of arithmetic precision on the BER, we show results using a half-precision floating-point format (fp16). Fig.~\ref{fig:fp16} reveals that the decentralized CD detector for $B_c=32$, $U=8$, $C=2$, and $B=64$, with fp16 has virtually no BER performance loss when compared to the double precision. This indicates that we can implement the proposed decentralized algorithms in hardware with low precision to reduce transfer bandwidth and complexity, without sacrificing the BER. The throughput results shown in Sec.~\ref{sec:datarate} support this claim.

\section{Multi-GPU implementations}

\subsection{System Architecture}
\label{sec:gpusys}
\begin{figure}[t]
\centering
\includegraphics[width=0.6\columnwidth]{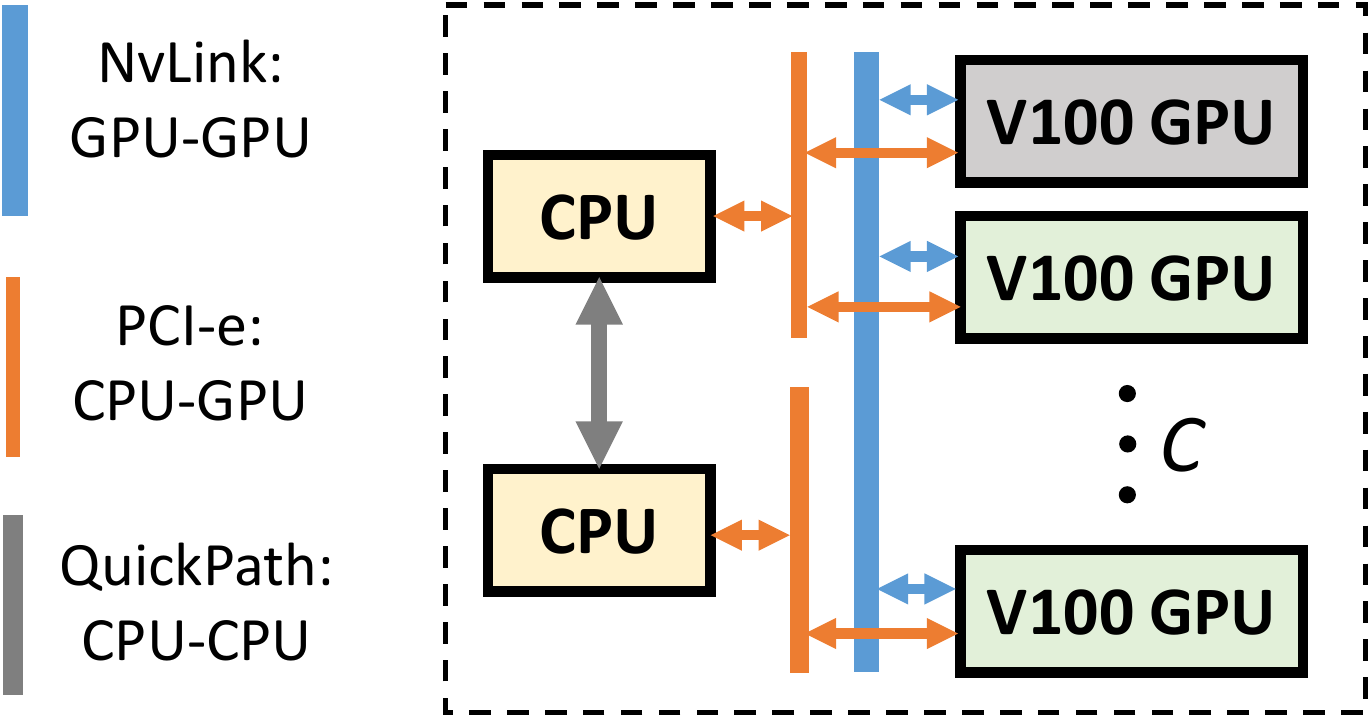}
\vspace{-0.1cm}
\caption{Overview of the multi-GPU system architecture. \revision{The top GPU, colored in gray, is the master GPU for data broadcasting or fusion.}}
\label{fig:gpuarch}
\end{figure}

We now present a multi-GPU implementation of the proposed decentralized CD data detector and precoder.
We implemented our designs on an Nvidia DGX-1~\cite{dgx1} multi-GPU system whose architecture is shown in Fig.~\ref{fig:gpuarch}. The system has two 20-core Intel Xeon E5-2698 v4 CPUs and eight Tesla V100 Volta GPUs with NvLink~\cite{nvlink} connection, a direct GPU-to-GPU communication link with 300\,GBps bi-directional bandwidth. Each V100 GPU consists of 5120 CUDA cores and 16\,GB high bandwidth memory (HBM). The designs are implemented with CUDA v9.2~\cite{cuda} for GPU acceleration of detection and precoding computations, and the NVIDIA Collective Communications Library (NCCL) v2.2~\cite{nccl}, which builds on the message passing interface (MPI) library for efficient data transfer among GPUs over NvLink. \revision{We implemented our designs with floating-point data types that are natively supported by the used GPUs.}

\subsection{Implementation Details}
\label{sec:impdetail}
We invoke $C$ MPI processes from the CPU controller, each process supervising the local CD detection or precoding computations on a GPU. The fusion of local detection results~$\hat{\bmx}_c$ in the uplink and broadcasting of the transmit signal $\bms$ in the downlink are realized by inter-process communication over the NvLink. Specifically, we use the \texttt{ncclReduce} function in NCCL for efficient data fusion and reduction, and use the \texttt{ncclBcast} function for fast broadcasting, both leveraging direct GPU-to-GPU memory copy for low latency. 

The local CD computations on each GPU are implemented by multi-threaded customized kernels. According to Algorithm~\ref{alg:dcd_det} and Algorithm~\ref{alg:dcd_pre}, the dominating operations of CD are vector operations instead of matrix operations. For vector scaling, addition, and subtraction operations, which have native parallelism, we straightforwardly generate multiple GPU threads to compute each element in parallel. Our algorithms also involve vector dot product and vector norm computations, which require aggregation, i.e., inter-thread communication, of element-wise product results in a pair of vectors. Here, we resort to the \emph{warp shuffle} technique to realize direct \emph{register-to-register} copy among threads within a \emph{warp}~\cite{cuda}, which significantly improves inter-thread communication efficiency compared to conventional solutions based on slower shared memories or global memories. \revision{To reduce high-latency global memory transactions, we combine multiple vector operations into a single kernel function, so that intermediate computation results are shared with fast on-chip memories within the kernel.} To fully exploit the GPU computing resources for high throughput, we process local detection or precoding workload for a large amount of OFDM subcarriers together with thousands of threads in parallel to keep high GPU occupancy.

To further improve throughput, we explored fp16 implementations supported by the latest CUDA release. Taking advantage of single-instruction multiple-data (SIMD) operations realized by CUDA fp16 \emph{intrinsic} functions~\cite{cuda}, in a single thread, we can simultaneously operate on two fp16 values packed inside a 32-bit \texttt{half2} type data, which leads to higher parallelism and computation efficiency. In addition, fp16 reduces the message size of inter-GPU data transfer to half of the fp32 message, decreasing transfer latency in our decentralized designs. 

\subsection{Data-Rate Performance}
\label{sec:datarate}
Table~\ref{tb:datarate} reports the latency and throughput of our CD data detector (DCD-D) and precoder (DCD-P) for various antenna configurations and both fp32 and fp16 implementations. We fix $U=8$ and $B_c=32$, and measure the performance of our decentralized designs on $C=\{4,8\}$ GPUs, which corresponds to $B=\{128, 256\}$. To characterize the performance of our implementations in a more practical setup, we assumed an LTE scenario in which we used batched processing workload with 1200 OFDM data subcarriers. For all antenna configurations, the data rate of fp16 based designs significantly outperforms the fp32 ones. If we scale up the total number of BS antennas by increasing $C$, then the throughput degrades only slightly; this demonstrates the scalability of our decentralized designs in systems with a large number of BS antennas.

Table~\ref{tb:perfcompare} compares the throughput performance of existing algorithms implemented on different computing fabrics, such as ASIC~\cite{BYICASSP13}, FPGA~\cite{WDCS2016}, and GPU~\cite{CJTSP2018,LJCS2018}. The proposed decentralized fp16 CD implementations achieve $2.5\times$ to $3.5\times$ improvements compared to existing fp32 fully-decentralized detectors and precoders in terms of per-user throughput where the throughput is normalized to 16-QAM\footnote{Scaling the modulation scheme to 16-QAM for architectures supporting higher-order constellations may penalize these designs as they were optimized for a different target rate; the penalty, however, is expected to be rather small.}.

\section{Conclusions}
\revision{We have presented decentralized CD-based data detection and precoding algorithms for massive MU-MIMO systems which mitigate the computation and interconnection bottlenecks in typical centralized designs. Our methods require lower complexity than existing methods without explicit matrix multiplication and inversions. We have demonstrated the hardware efficiency and design scalability with fp32 and fp16 multi-GPU implementations, and have  realized a $3\times$ speed up on per-user throughput compared to previous decentralized detectors and precoders. We conclude by noting that our current GPU implementations serve as a fast prototyping solution of our algorithms, while the power dissipation of $C$ Tesla V100 GPUs can be as high as $C\times 300$ W. To reduce power, one can resort to multi-FPGA and multi-ASIC designs; an analysis of such more efficient solutions is left for future work.}

%

\end{spacing}


\balance
\bibliographystyle{IEEEtran}
\bibliography{bib/VIPabbrv,bib/confs-jrnls,bib/publishers,bib/VIP_180730}

\balance

\end{document}